# Fuzzy Model Tree For Early Effort Estimation

Mohammad Azzeh
Department of Software Engineering
Applied Science University
Amman, Jordan
m.y.azzeh@asu.edu.jo

Ali Bou Nassif
Department of Electrical & Computer Engineering
Western University
London, Ontario, Canada, N6A 5B9
abounas@uwo.ca

*Abstract*— Use Case Points (UCP) is a well-known method to estimate the project size, based on Use Case diagram, at early phases of software development. Although the Use Case diagram is widely accepted as a de-facto model for analyzing object oriented software requirements over the world, UCP method did not take sufficient amount of attention because, as yet, there is no consensus on how to produce software effort from UCP. This paper aims to study the potential of using Fuzzy Model Tree to derive effort estimates based on UCP size measure using a dataset collected for that purpose. The proposed approach has been validated against Treeboost model, Multiple Linear Regression and classical effort estimation based on the UCP model. The obtained results are promising and show better performance than those obtained by classical UCP, Multiple Linear Regression and slightly better than those obtained by Tree boost model.

I.  INTRODUCTION

One long standing question in software engineering is how to accurately predict the required software development effort at the early stage of software development [1, 2]. The historical records of some large software projects show inferior quality with respect to time and cost estimation, and the consequences to business issues can be enormously damaging. Lynch [3] showed that over 50% of large software projects significantly overrun their estimates (with an error percentage that can vary from 100% to 200%) and 15% of them were never completed due to the gross misestimating of development effort. Both underestimation and overestimation can cause severe problems to software projects such that underestimation leads to understaffing and consequentially takes longer to deliver project than necessary, whereas overestimation may lead to miss opportunities to offer funds for other projects in future [8].

Software effort estimation at the very early stages of software development is imperative. The project manager needs to provide an initial estimate of the size and effort required to develop the software product [8]. These estimates are very important for feasibility study and project bidding, but for some reasons these early estimates are only guesses, with inherent uncertainty and risks. On the other hand, project size provides a general sense of how large a software project may be; hence, it gives an indicator of how many resources this project really needs to be developed. In spite of existence of many size estimation techniques, Line of Code (LOC) and Function Points remains the most used methods in many software cost estimation models. The LOC metric [4] is criticised because it is programming language dependent and it is not recommended to be used in the early phases because the source code is not yet available. Function points (FP) [5] appears as well suited method to measure early project size because it measures the amount of functionalities of the system that are available during the requirements analysis [16]. A significant major problem of FP is being restricted to business application and management information system [16]. Moreover, counting FP is tedious and requires special care. To conclude, both LOC and FP are incompatible with object oriented software development.

Therefore the Use Case Points (UCP) [6] came out as new solution to support early estimation and object oriented software development. UCP is a relatively new and rather simple to use. The philosophy behind a UCP is quite similar to that of function points and basically dependent on the Use Case diagram. Experienced software estimators are required to translate the set of requirements into their likely number of use cases, actors and scenarios [16]. Use Case Points are calculated through a systematic procedure and its accuracy level depends on the degree of the use case diagram details. However, UCP is best suited for OO programming languages and can help the project manager to measure software application size at early stage of software development. Although Use Case diagram is a de-facto analysing model, the UCP is not widely accepted in software industries because there is no consensus on how to translate the derived UCP into corresponding effort.This paper aims to study the use of Fuzzy Model Tree to produce the required efforts from software projects measured by UCP. The Model Tree (MT) [12, 13] is a special type of decision tree model developed for the task of non-linear regression. However, the main difference between MT and regression trees is that the leaves of regression trees present numerical values, whereas the leaves of a MT have regression functions. Fuzzy sets has been integrated with MT to partition training data set based on their membership values in each cluster as explained in section III .

The present paper is structured as follows: Section II gives background about the employed techniques. Section III presents the proposed approach. Section IV discusses the methodology of this research. Section V presents the obtained results, and finally Section VI ends with the conclusions.

## II. BACKGROUND

### A. Use Case Points

The use case point (UCP) [6] size estimation model was first described by Gustav Karner in 1993 [6]. The process of calculating UCP size measure is briefly described in the following four steps:
1) From use case diagram, identify and classify types of actors and use cases into *simple*, *average* and *complex*. Classifying actors and use cases helps in weighting size estimation according to the system complexity. The resulted values are Unadjusted Weighted Actors (UWA), and Unadjusted Use Case Counts (UUC)
2) Calculating Unadjusted Use Case Points (UUCP) by adding UWA and UUC.
3) Compute Technical and Environmental Factors from pre-existed table.
4) The Adjusted Use Case Points UCP is calculated by multiplying unadjusted use case count UUCP by Technical and Environmental factors.

### B. Model Tree

The Model Tree (MT) [12, 13] is another form of regression trees where the difference lies in producing prediction. The leaves of regression trees present average of dependent values of all instances in the leaf node, whereas the leaves of MT have regression functions as shown in Fig. 1. The MT is constructed through an iterative process known as binary recursive partitioning method such that training dataset is split into number of partitions, and then splitting it up further on each of the branches. In this paper we used M5P algorithm which has three stages [12]: First, a decision-tree induction algorithm is used to build a tree, but instead of maximizing the information gain at each inner node, a splitting criterion is used that minimizes the intra-subset variation in the class values down each branch. The splitting procedure in M5P stops if the class values of all instances that reach a node vary very slightly, or only a few instances remain. Second, the tree is pruned back from each leaf. When pruning an inner node is turned into a leaf with a regression plane. Third, a smoothing procedure is applied to avoid sharp discontinuities between adjacent linear models at the leaves of the pruned tree. This procedure combines the leaf model prediction with each node along the path back to the root, smoothing it at each of these nodes by combining it with the value predicted by the linear model for that node.

### C. Fuzzy Model

Fuzzy logic and sets provide a representation scheme and mathematical operations for dealing with uncertain, imprecise and vague concepts. Fuzzy logic is a combination of a set of logical expressions with Fuzzy sets. Zadeh [14] defined the meaning of the membership for Fuzzy sets to be a continuous number between zero and one. Each Fuzzy set is described by a membership function such as Triangle, Trapezoidal, Gaussian, etc., which assigns a membership value between 0 and 1 for each real point on universe of discourse. Fuzzy models can be constructed by one of two ways either by expert knowledge or using algorithms. The former, uses the experience that is formed in if-then-rules expressions where parameters and memberships are tuned using input and output data. The latter uses algorithms such as Fuzzy C-means (FCM) [15] to create membership functions. For instance, the Fuzzy model in this paper was constructed based on the second approach.

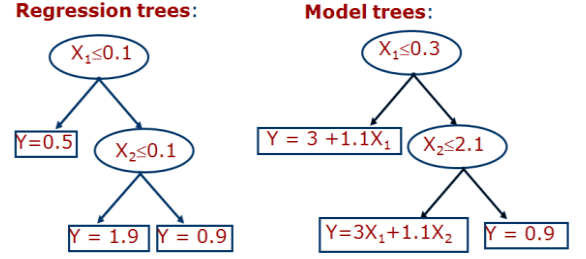

Fig. 1. Difference between regression trees and Model Trees.

## III. THE PROPOSED FUZZY MODEL TREE

The proposed approach has integrated both Fuzzy Modelling with Model Tree to produce a hybrid approach called Fuzzy Model Tree. The proposed approach starts with using Fuzzy C-means to partition training dataset into several clusters where each instance may belong to different clusters with different membership values. The obtained membership values are then used to construct Fuzzy model using *genefis3* MATLAB® function. *genfis3* is a function used to construct a Fuzzy model based on the concept of Fuzzy clustering. The process of Fuzzy model construction can be understood by the following simple illustration. Suppose there are *N* data samples that are described by 3 dimensional features (FA. FB and FC) as shown in Fig. 2(a) which are clustered using FCM algorithm into 3 Fuzzy clusters as shown in Fig. 2(b). These Fuzzy clusters are then used to construct their corresponding Fuzzy sets on each universe of discourse as shown in Fig. 2(c), 2 (d) and 2(e).

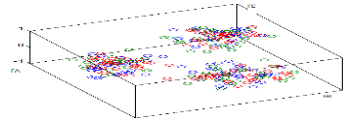 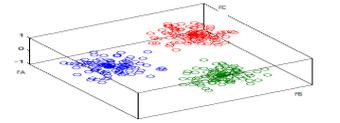

Fig. 2(a). Training dataset     Fig. 2(b). Clustered dataset

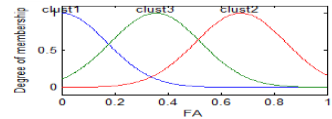 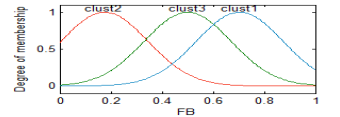

Fig. 2(c). Membership functions for feature FA.     Fig. 2(d). Membership functions for feature FB.

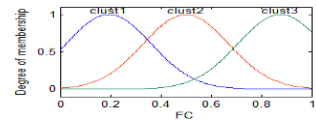

Fig. 2(e). Membership functions for feature FC.

Then we compute membership values for each training project feature in all clusters as shown in Table I where $\mu_{ij}(C_k)$ is the membership value of project *i* at feature *j* in cluster $C_k$. The obtained membership matrix with the actual training efforts forms the input for Model Tree. The all instances in the leaf nodes are then used to construct linear regression models based on the actual feature values of training projects (i.e. not the membership values).

TABLE I. MEMBERSHIP VALUES FOR EACH PROJECT

| # | $Feature_1$ | | | | $Feature_2$ | | | | $Feature_j$ | | |
|---|---|---|---|---|---|---|---|---|---|---|---|
| 1 | $\mu_{11}(C_1)$ | $\mu_{11}(C_2)$ | … | $\mu_{11}(C_k)$ | $\mu_{12}(C_1)$ | $\mu_{12}(C_2)$ | … | $\mu_{12}(C_K)$ | … | $\mu_{1j}(C_k)$ | … |
| 2 | $\mu_{21}(C_1)$ | $\mu_{21}(C_2)$ | … | $\mu_{21}(C_k)$ | $\mu_{22}(C_1)$ | $\mu_{22}(C_2)$ | … | $\mu_{22}(C_K)$ | … | $\mu_{2j}(C_k)$ | … |
| … | …. | …. | … | …. | … | … | … | … | … | … | … |
| i | $\mu_{i1}(C_1)$ | $\mu_{i1}(C_2)$ | …. | $\mu_{i1}(C_k)$ | $\mu_{i2}(C_1)$ | $\mu_{i2}(C_2)$ | … | $\mu_{i2}(C_k)$ | … | $\mu_{ij}(C_k)$ | … |
| … | …. | …. | … | …. | … | … | … | … | … | … | … |

For testing purposes, we first compute membership values for each testing project in the constructed Fuzzy model in the training phase. Then, the obtained membership values are used as input to the obtained Model Tree.

IV. METHODOLOGY

A. Dataset Description

The dataset employed in this paper has been collected based on UCP sizing technique, using a questionnaire conducted by [1]. The collected dataset is described by four features: the size in UCP based on the UCP model, productivity, complexity and actual effort. The productivity and complexity features are described in [1]. The dataset contains 84 projects: 58 industrial projects and 26 educational projects collected from three main sources. Table II shows the characteristics of these datasets.

TABLE II. UCP DATA CHARACTERISTICS

| Source | Ind1 | Ind2 | Edu |
|---|---|---|---|
| Min Effort (PH) | 4,648 | 570 | 850 |
| Max Effort (PH) | 129,35 | 224,890 | 2380 |
| Mean Effort | 36,849 | 20,573 | 1,689 |
| Standard Deviation (Effort) | 39,350 | 47,327 | 496 |
| Skewness (Effort) | 1.37 | 3.26 | -0.24 |

B. Evaluation Criteria

The prediction accuracy of different techniques is assessed using *MMRE*, *MdMRE*, *pred(0.25)* and *pred(0.5)*.
*MMRE* computes mean of the absolute percentage of error between actual ($x_i$) and predicted ($\hat{x}_i$) project effort values as shown in Eq. 1 and 2. *MdMRE* computes median of MREs as shown in Eq. 3.

$$MRE_i = \frac{|x_i - \hat{x}_i|}{x_i} \quad (1)$$

$$MMRE = \frac{1}{N}\sum_{i=1}^{N} MRE_i \quad (2)$$

$$MdMRE = \underset{i}{median}(MRE_i) \quad (3)$$

*pred(0.25) and pred(0.5)* are used as complementary criterion to count the percentage of MREs that fall within less than 0.25 and 0.5 respectively of the actual values as shown in Eq. 4.

$$pred(l) = \frac{100}{N} \times \sum_{i=1}^{N} \begin{cases} 1 & if \ MRE_i \leq l \\ 0 & otherwise \end{cases} \quad (4)$$

The Boxplot of absolute residuals is also used to compare between different models. The length of Boxplot from lower tail to upper tail shows the spread of the distribution. The length of box represents the range that contains 50% of observations. The position of median inside the box and length of Boxplot indicates the skewness of distribution. A Boxplot with a small box and long tails represents a very peaked distribution while a Boxplot with long box represents a flatter distribution. In addition to that we used *win-tie-loss* algorithm [17] to compare the performance of FMT to other model as shown in Fig. 3. To do so, we first check if two methods $M_i$; $M_j$ are statistically different according to the Wilcoxon test; otherwise we increase $tie_i$ and $tie_j$. If the distributions are statistically different, we update $win_i$; $win_j$ and $loss_i$; $loss_j$, after checking which one is better according to the performance measure at hand *E*. The performance measures used here are *MRE, MMRE, (MdMRE), pred(0.5)* and *pred(0.25)*.

```
1   win_i=0,tie_i=0,loss_i=0
2   win_j=0,tie_j=0;loss_j=0
3   if WILCOXON(MRE(M_i), MRE(M_j), 95) says they are the
4   same then
5          tie_i = tie_i + 1;
6          tie_j = tie_j + 1;
7   else
8          if better(E(M_i), E(M_j)) then
9                 win_i = win_i + 1
10                loss_j = loss_j + 1
11         else
12                win_j = win_j + 1
13                loss_i = loss_i + 1
14         end if
    end if
```

Fig. 3. Pseudo code for *win-tie-loss* calculation between method $Mi$ and $Mj$ based on performance measure *E* [17].

## C. Empirical Evaluation

The proposed FMT technique was compared to three techniques that were evaluated previously on the same dataset. These techniques are: Treeboost model [1], Multiple Linear Regression (MLR) and classical UCP technique (UCP) [6]. The FMT, Treeboost model and MLR have been built using 59 projects and the remaining projects were used for testing.

### 1) Treeboost Model:

The Treeboost model has been previously used by A.B. Nassif et al. [1] to improve the accuracy of software effort estimation based on UCP size measure. The Treeboost model is also called Stochastic Gradient Boosting (SGB) [9, 10]. Boosting is a method to increase the accuracy of a predictive function by applying the function frequently in a series and combining the output of each function. The main difference between the Treeboost model and a single decision tree is that the Treeboost model consists of a series of trees [7]. The main limitation of the Treeboost is that it acts like a black box (similar to some neural network models) and cannot represent a big picture of the problem as a single decision tree does. The Treeboost algorithm is described in Eq. 5:

$$F(x) = F_o + A1 \times T1(x) + A2 \times T2(x) + \cdots + Am \times Tm(x) \quad (5)$$

Where $F(x)$ is the predicted target, $F_0$ is the starting value, $x$ is a vector which represents the pseudo-residuals, $T1(x)$ is the first tree of the series that fits the pseudo-residuals (as defined below) and $A1$, $A2$, etc. are coefficients of the tree nodes. The full description of the Treeboost model can be found in [1]. In this paper we used the same configuration parameters that have been used by Nassif et al. [1] which are: (number of trees=1000, Huber Quantile Cutoff=0.95, Shrinkage Factor=0.1, Stochastic Factor= 0.5, Influence Trimming Factor=0.01).

### 2) Regression model:

The MLR model has been built over 59 training projects, but before that we made sure that all assumptions related to using MLR are not violated [11]. For example, skewed numerical variables need to be transformed such that they resemble more closely a normal distribution. The applied normality test suggests that "Effort" and "Size" were not normally distributed, so "$ln$(effort)" and "$ln$(size)" were used instead of "Effort" and "Size". The logarithmic transformation ensures that the resulting model goes through the origin on the raw data scale. It also caters for both linear and non-linear relationships between size and effort. The resulting MLR mode is shown in Eq. 6.

$$ln(Effort) = 1.8 + 1.24 \times ln(Size) + 0.007 \times Productivity + \quad (6)$$
$$0.12 \times Complexity$$

Where Effort is measured in person-hours and Size in UCP. The adjusted $R^2$ of the MLR is 0.8 which suggests that the model was fairly good with 80% of the variation in effort being explained by variation in size, productivity and complexity. The statistical significance analysis showed that $ln(size)$ and complexity variables are statistically significant at the 95% confidence level. We also measured the Variance Inflation Factor (VIF) of each independent variable to see if the multicollinearity issue (when one independent variable has a relationship with other independent variables) exists. We found that the highest VIF factor is for the variable "$ln$(Size)" which is 1.03. This indicates that the multicollinearity issue does not exit (VIF is less than 4).

### 3) Classical UCP

The classical way to predict effort is to multiply the obtained UCP with the productivity ratio. Since in earlier model there was no historical projects collected on UCP it was very hard to compute productivity ratio, therefore a figure between 15 and 30 was suggested by industry experts as demonstrated in [18]. A typical value is 20 that means one UCP requires roughly 20 person-hour as shown in Eq. 7.

$$Effort = UCP \times 20 \quad (7)$$

## V. RESULTS AND DISCUSSION

This section presents the obtained results when applying different models. For validation purposes we used 59 projects as training and 25 projects as testing. The performance figures of empirical validation are presented in Table III. Upon results analysis of the empirical validation, FMT shows better estimation accuracy than other models in terms of all evaluation criteria, but Treeboost model produced remarkable accuracy in terms of *MdMRE*. The MLR produced slightly the worst results, suggesting that the characteristic of this dataset is not linear so the non-linear regression model would perform better. Also the results show that at least 96% of generated predictions by FMT have MRE values less than 50%. Furthermore, the difference between the best and worst accuracy is remarkable and an indication of the performance of FMT.

TABLE III. PERFORMANCE FIGURES

| Model | MMRE | MdMRE | Pred(0.25) | Pred(0.5) |
|---|---|---|---|---|
| FMT | 21.8 | 19.4 | 64.0% | 96.0% |
| Tree boost | 29.0 | 14.0 | 64.0% | 88.0% |
| MLR | 44.0 | 44.0 | 8.0% | 60.0% |
| UCP | 38.0 | 40.0 | 40.0% | 64.0% |

The boxplot of absolute residuals as shown in Fig. 4 provides a better insight on the effectiveness of Non-linear models prediction models such as FMT and Treeboost. Figure 1 shows that classical UCP produced the worst estimates with extreme absolute residual values. This problem may be caused by using the same productivity weight for all projects irrespective of their size and environment. However, the box of FMT overlays the lower tail which shows that the absolute residuals are skewed towards the minimum value and also presents accurate estimation than other three models. The range of absolute residuals of FMT is much smaller than absolute residuals of UCP and MLR which also presents smaller variance. The median of FMT and Treeboost are smaller than the median of other models which revealed that at least half of the predictions are more accurate than those generated by MLR and UCP.

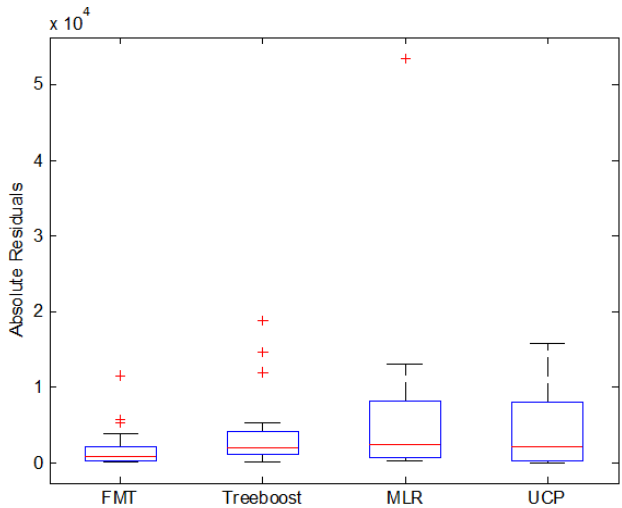
Fig. 4. Boxplot of absolute residuals

To identify top model over the employed dataset, we run *win-tie-loss* algorithm. This algorithm ranks different methods based on comparison between them in terms of some performance measures over the employed dataset. The overall results of *win-tie-loss* are recorded in Table IV. However, there is reasonable believe that using FMT has never been outperformed by other models. Indeed, this confirms the significant improvement brought to the early effort estimation. From these results we can notice that the number of *win-loss* suggests that FMT is the best performer with 10 *wins* and one *loss*. However, the Treeboost still produces remarkable results to FMT with *win-loss*=7 which shows the potential of both models to produce more accurate estimate at early stages of software development.

TABLE IV. WIN-TIE-LOSS RESULTS

| Method | Win | Tie | loss | Win-loss | Rank# |
|---|---|---|---|---|---|
| FMT | 10 | 1 | 1 | 9 | 1 |
| Tree boost | 9 | 1 | 2 | 7 | 2 |
| MLR | 0 | 1 | 8 | -8 | 3 |
| UCP | 0 | 1 | 8 | -8 | 3 |

## VI. CONCLUSIONS

Software effort estimation is recognized as a regression problem and machine learning methods such as Regression Tree, Model Tree (MT), Support Vector Machine, Radial Basis Functions, etc. are more capable of handling noisy datasets than statistical based regression models that focus on the correlation between variables. This paper proposed a new Fuzzy Model Tree for improving early software effort estimation accuracy based on the use of UCP. The use of FMT enabled us to classify instances based on their membership values in the derived Fuzzy clusters. The FMT was built over 59 training project and evaluated using 25 projects. The results obtained for FMT are encouraging and better than previous models. However, publication of raw results is still important so further research is necessary to investigate the real implications of the proposed model on more industrial projects rather than educational projects.

ACKNOWLEDGMENT

The authors are grateful to the Applied Science University, Amman, Jordan, for the financial support granted to cover the publication fee of this research article.